\documentstyle[epsfig,pre,aps]{revtex}

\begin{document}
\title{Combining quantum and classical density functional theory
for ion-electron mixtures.}

 \author{A.A. Louis} \address{ Department of Chemistry, Cambridge
University, Lensfield Rd, Cambridge CB2 1EW, UK}

\author{H. Xu} \address{D\'epartement de Physique des Mat\'eriaux (UMR
5586 du CNRS), Universit\'e Claude Bernard-Lyon I, 43 Boulevard du 11
Novembre 1918, 69622 Villeurbanne Cedex, France.}

\author{J.A. Anta} \address{Departamento de Ciencias Ambientales,
Universidad Pablo de Olavide, 
Crta de Utrera Km.1, 41013 Sevilla, Spain
}

 \date{\today} \maketitle
\begin{abstract}
We combine techniques from quantum and from classical density
functional theory (DFT) to describe electron-ion mixtures. For
homogeneous systems, we show how to calculate ion-ion and ion-electron
correlation functions within Chihara's quantum hypernetted chain
approximation, which we derive within a DFT formulation.  We also
sketch out how to apply the DFT formulation to inhomogeneous
electron-ion mixtures, and use this to study the electron distribution
at the liquid-solid interface of Al.
\end{abstract}


\section{Introduction}

Density functional theory (DFT) has proven itself a remarkably
successful tool in condensed matter physics\cite{Kohn}. The
foundations were laid in 1964, when Hohenberg and Kohn
(HK)\cite{Hohe64} proved that the ground state energy of any quantum
mechanical system could be described as a functional of the one-body
density only.  Subsequently, Kohn and Sham\cite{Kohn65} developed an
orbital based method which could be applied to electronic systems, and
Mermin\cite{Merm65} extended the HK proof to finite temperatures,
opening up the possibility of using DFT to calculate the free-energy
of a statistical mechanical system.  Since then many different
practical methods have been developed to apply DFT to electronic
problems, with countless applications in condensed matter physics,
chemistry, and biology\cite{Gros95}.  Examples of electronic structure
techniques relevant to this paper are the {\em ab initio} molecular
dynamics (AIMD) method of Car and Parrinello\cite{Car85} and the
orbital free {\em ab initio} molecular dynamics (OF-AIMD) scheme of
Madden and co-workers\cite{Pear93}. In a parallel development, the
finite temperature version of DFT has been widely used to study
classical systems, with myriad applications to phase-transitions and
the theory of fluids (in its broadest sense), see
e.g.\cite{Baus90,Evan92} for reviews.

However, the classical and quantum versions of DFT have rarely been
combined -- the two fields have largely developed independently.  In
this paper we will attempt to bring together these two strands of DFT
into one unified formalism, with the aim of applying them to liquid
metals. These are ideal systems for which to test a mixed
quantum-classical DFT approach since they can be viewed as binary
mixtures of classical ions and quantum electrons.

To begin, we first define a two-component free-energy functional
\begin{equation}\label{eq1.1}
F[\rho_I,\rho_e] = F_I^{id}[\rho_I] + F_e^{id}[\rho_e] + 
F^{ex}[\rho_I,\rho_e],
\end{equation}
split into the usual way between the ideal free energies of the ions,
$F_I^{id}$, and the electrons, $F_e^{id}$, and the excess free energy
$F^{ex}$, which is a functional of both one-body densities
$\rho_I({\bf r})$ and $\rho_e({\bf r})$ .
  This
unique intrinsic free-energy functional functional obeys the
variational principle:
\begin{equation}\label{eq1.2}
\frac{\delta F[\rho_I,\rho_e]}{\delta \rho_\alpha({\bf r})} + \phi_\alpha({\bf r}) = 0,
\end{equation}
where the $\phi_\alpha({\bf r})$ are the external potentials which
uniquely induce the one-body densities $\rho_\alpha({\bf r})$.  This
principle is subject to the constraint that the integral of
$\rho_\alpha({\bf r}) = N_\alpha$, the total number of particles of
species $\alpha = \{e,I\}$.
We will apply this free-energy functional both to homogeneous and
to inhomogeneous liquid metals.

  In section~\ref{section2} we use the DFT formulation to derive the
quantum Ornstein Zernike(QOZ) equations. From these we can derive a
set of integral equations, called the quantum hypernetted chain
approximation(QHNC) by Chihara\cite{Chih78}, which self-consistently
solves for the ion-ion and ion-electron pair correlation functions in
a homogeneous liquid metal.  The great simplification of the QHNC
approach is that the original many-centre electronic problem is
reduced to an effective one-centre problem, which is much easier to
solve.  As a specific application, we study the ion-electron radial
distribution function for a set of simple metals.

In section~\ref{section3} we extend this DFT formulation to
inhomogeneous fluids.  By using a simple version of this functional,
we compare the electron density profile at the liquid-solid interface
of Al to the recent simulations using OF-AIMD\cite{Jess00}. We end
with some conclusions.

\section{Applying two-component DFT to homogeneous liquid metals}\label{section2}

\subsection{Quantum Ornstein Zernike equations}
DFT provides a nicely unified route to the correlation functions of
homogeneous fluids.  To begin, we first show how to derive the QOZ
relations by defining the direct correlation function as the second
functional derivative of the excess free-energy defined in
Eq.~(\ref{eq1.1}):
\begin{equation}\label{eq2.1}
\frac{1}{\beta} C_{\alpha \beta}({\bf r},{\bf r'}) = \frac{-\delta^2
F^{ex}}{\delta \rho_\alpha({\bf r}) \delta \rho_\beta({\bf r'})} =
(\chi_{\alpha \beta})^{-1} - (\chi^0_{\alpha \beta})^{-1},
\end{equation}
where $(\chi_{\alpha \beta})^{-1}$ is the inverse susceptibility
matrix of the full system, and $(\chi^0_{\alpha \beta})^{-1}$ is the
inverse susceptibility matrix of the ideal system.  These
relationships between direct correlation functions and inverse
susceptibilities are called the QOZ equations, and they hold for an
arbitrary number of components.  If any components are classical, then
the susceptibilities can be connected to the pair-correlations through
the fluctuation-dissipation theorem\cite{Kubo64}:
\begin{equation}\label{eq2.2}
\lim_{\hbar \rightarrow 0}\chi_{\alpha \beta}(k,0) =
-\beta(\rho_\alpha^0 \rho_\beta^0)^{1/2} S_{\alpha \beta}(k),
\end{equation} where the $S_{\alpha \beta}(k)$ are the  structure
factors, defined in the usual way\cite{Ashc78,Hans86}, and we have
taken the homogeneous limit and the $\rho_alpha^0$ are the homogeneous
limits of the densities $\rho_\alpha({\bf r})$.  Specializing to the
homogeneous limit for two components, with one component being
classical (the case at hand), the QOZ equations reduce to a set of
relationships between the direct correlation functions and the the
ion-ion and ion-electron pair correlation functions as well as the
electron-electron susceptibilities. The latter are still
quantum-mechanical, and cannot be simply connected to
electron-electron pair correlations.

\subsection{Quantum Hypernetted Chain Approximation}

To make further progress, one needs a way of solving the QOZ relations
for a liquid metal. We sketch the derivation here, for more details we
refer to reference\cite{Anta00} and to the original work of
Chihara\cite{Chih85,Chih89}. As a first step, we use a quantum version
of the Percus trick\cite{Perc62} to relate the {\em homogeneous}
two-body pair-correlation functions to the one-body {\em
inhomogeneous} density around a single classical particle fixed at the
origin.  For the ion-electron pair-correlation function we fix an ion
at the origin to find:
\begin{equation}\label{eq2.3}
g_{Ie}({\bf 0,r}) = \frac{\rho_e({\bf r}|I)}{\rho_e^0},
\end{equation}
where $\rho_e({\bf r}|I)$ is the (interacting) valence electron
density\cite{core-electrons}.  A similar relationship can be found for
the ion-ion pair-correlation function, but this trick cannot be used for the
electron-electron pair-correlation function, since one cannot ``fix''
an electron at one position.  The next step is to solve for the
interacting one-body electron-density.  To do this, one can use the
Kohn-Sham trick\cite{Kohn65}, namely that there exists a 
single-particle external potential $v^{\text{eff}}({\bf r})$ which
will induce in a {\em non-interacting} system the same one-particle
density $\rho({\bf r})$ as is found in the full {\em interacting}
system.  This external effective potential, felt by the non-interacting
electrons or non-interacting ions, follows from the Euler equations. If
one further makes a functional Taylor expansion around the equilibrium
homogeneous densities, then the following effective ion-ion and
ion-electron potentials result:
\begin{equation}\label{eq2.4}
v_{\alpha I}^{\text{eff}}(r) = v_{\alpha I}(r)
   -\frac{1}{\beta}\sum_\gamma\rho_\gamma\int 
C_{\alpha\gamma}(|{\bf r}-{\bf r}^\prime|)
             h_{\gamma I}(r)d{\bf r}^\prime
+\frac{1}{\beta}B_{\alpha I}(r),
\end{equation}
where the $C_{\alpha \gamma}(r)$ are the homogeneous limits of the
direct correlation functions defined in Eq.~(\ref{eq2.1}), and the
Percus trick was used to rewrite $(\rho_\gamma({\bf r}|v_{\gamma I}) -
\rho_\gamma^0)$ in terms of the correlation functions $h_{\gamma I}(r)$
= $g_{\gamma I}(r) -1$.  The remaining higher order terms are lumped
into the so-called {\em bridge functions} $B_{\alpha \beta}(r)$, well
known in the theory of liquids\cite{Rose79,Hans86}.  Again, we note that no
such potential can be derived for the electron -electron correlations.
With that caveat in mind, our formulation, derived from a DFT
approach, is still in principle exact.  

To make progress, however, some approximations must be made, in
particular to treat the electron-electron correlations.  We follow
Chihara, who made a series of such approximations to derive 
the QHNC equations\cite{Chih78}. 
The most important one is:

\begin{enumerate}

\item{\em The valence electron correlations are treated in the jellium
approximation}. I.e.\ the effect of the ion-ion and ion-electron
correlations on the electron-electron correlations are ignored, and
the electron-electron direct-correlation function is written as:
\begin{equation}\label{eq2.5} C_{ee}(k) = -\beta
v_{ee}(k)[1-G_{ee}^{jell}(k;\rho_e^0)],
\end{equation}
where $G_{ee}$ is the local field factor for the electron
gas\cite{Hafn87,Ichi81}.  This has the great advantage that the QOZ
relations~(\ref{eq2.1}) now reduce to only two coupled equations for
the ion-ion and the ion-electron correlations.  While this is the key
approximation that makes the QOZ relations tractable, it is also
the most important and uncontrolled approximation in the QHNC scheme.
The fact that the effective pair potentials derived from a linear
response scheme\cite{Ashc78,Hafn87} are very sensitive to the exact
form of the local field factor also testifies to the importance of the
approximations in Eq.~(\ref{eq2.5}) to the physics of a liquid metal.

QHNC also uses an implicit separation of the valence electrons from
the core electrons. This approximation is closely related to the
jellium approximation discussed above, and is only good when the core
and valence electron energy levels are well separated so that the
ionic correlations don't have a significant effect on the core states;
it is likely to break down near resonances.

The other approximations used in QHNC are much better understood, and
are often  encountered in either their electronic structure\cite{Hafn87}
or liquid state theory context\cite{Hans86}.  In descending order of
importance they are (For a more detailed discussion see
ref.~\cite{Anta00}.):

\item{\em The local density approximation (LDA) is used for the
one-centre ion-electron problem}. This approximation is well known and
understood in electronic-structure theory\cite{Hafn87}.  Although in
principle this approximation is similar to approximation 1 for
$C_{ee}(Q)$, since the exchange and correlation energy of an
inhomogeneous electron gas are approximated by the local energy
density of jellium, there are a number of reasons to believe the LDA
is rather accurate\cite{Gros95}.  Various improvements to the LDA
exist, but these are not thought to be very important for the simple
metals we study.

\item{The ion-electron bridge function $B_{Ie}(r)$ is set to 0}, which
is related to the hypernetted-chain approximation, well known in
liquid state theory \cite{Hans86}.

\item{The ion-ion bridge function $B_{II}(r)$ is approximated by the
bridge-function of a hard-sphere reference state}.  This is called the
RHNC or MHNC approximation in liquid state theory\cite{Hans86,Rose79},
and is generally very accurate.

\item{The bare ion-ion potential is taken to be purely
Coulombic}. Again, this is an approximation whose limitations are well
understood\cite{Hafn87}.
\end{enumerate}

\begin{figure}
\begin{center}
\epsfig{figure=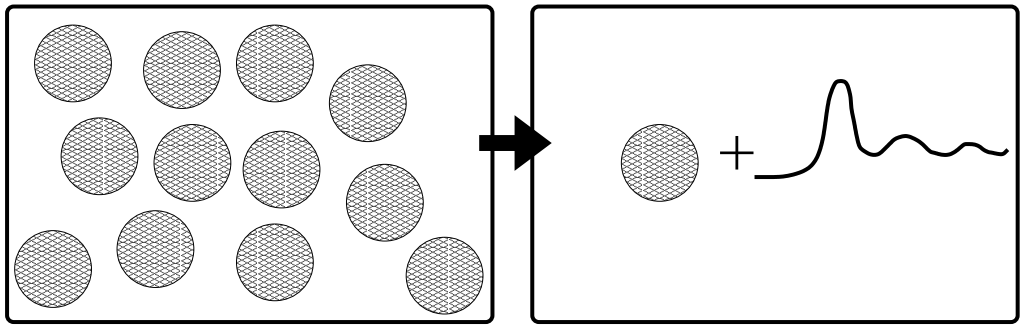,width=12cm,angle=-90} 
\begin{minipage}{14cm}
\caption{ Schematic picture of the QHNC approximation:  The original
many-centre problem is reduced to an effective one-centre problem.
}\label{fig:qhnc-schematic}
\end{minipage}
\end{center}
\end{figure}

In summary then, the QHNC approximation achieves the following very
important simplification for the electronic problem: the original
many-centre electronic problem has been reduced to an effective
one-centre problem by replacing the direct effect of the ions with an
effective external potential, given by Eq.~(\ref{eq2.4}), that depends
self-consistently on the ion-ion correlations. This is depicted
schematically in Fig.~\ref{fig:qhnc-schematic}.  The main advantages
are that the ion-ion and ion-electron correlations emerge naturally
and on the same footing and that the calculations are much more rapid
than full {\em ab initio} simulations.  By using DFT to derive the
QHNC equations, their origin and meaning becomes more transparent.
For details on the (non-trivial) numerical implementation of the QHNC
we refer to refs.~\cite{Anta00,Chih85,Chih89}.

\subsection{Applications of the QHNC to ion-electron correlations}

\begin{figure}
\begin{center}
\epsfig{figure=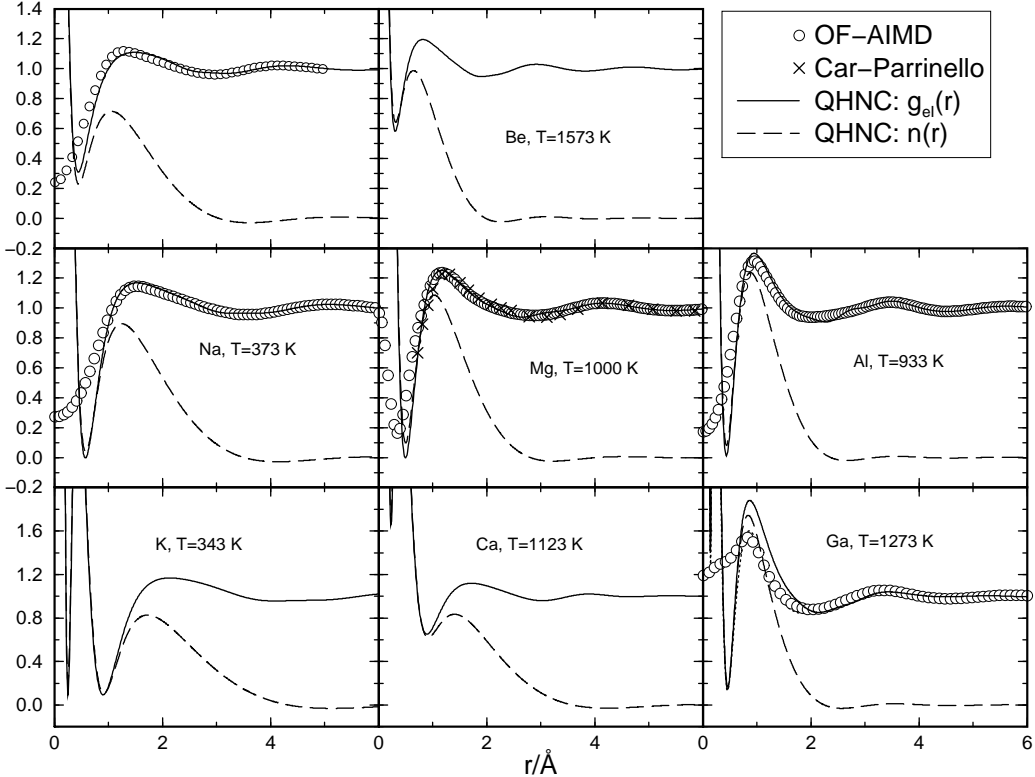,width=12cm,angle=-90} \vglue 0.1cm
\begin{minipage}{14cm}
\caption{The ion-electron radial distribution functions as obtained
from the QHNC approximation (solid lines),
OF-AIMD\protect\cite{Anta98,Anta99} (open circles) and Car-Parrinello
AIMD\protect\cite{deWi95}(crosses). The dashed lines represent the
pseudo-atom density $n(r)/\rho_e^0$.  \protect\label{fig:gei}}
\end{minipage}
\end{center}
\end{figure}

The ion-electron radial distribution function, defined as the
conditional probability of finding a valence electron a distance $r$
away, given that there is an ion at the origin, can be written as:
\begin{equation}\label{eq3.6}
\rho_e^0 g_{Ie}(r) = n(r) + \rho_e^0 \int_V n(r-r') g_{II}(r')
d{\bf r}',
\end{equation}where $n(r)$ is the so-called 
``pseudo-atom'' density, which, when superimposed according the
ion-ion radial-distribution function $g_{II}(r)$, gives the correct
value of the total electron-density.  The ion-electron radial
distribution function and the related pseudo-atom density, calculated
with the QHNC approach, are shown in Fig.~\ref{fig:gei} for a series
of simple metals.  At short distances the probability of finding an
electron a distance $r$ away is equal to the pseudo-atom density, but
further away the effects of the other surrounding pseudo-atoms kick
in.  Note that the QHNC is an all-electron calculation; the
oscillations near the core are correctly
reproduced\cite{core-electrons}, in contrast to the traditional AIMD
approaches, which rely on pseudopotentials.  The ion-electron
correlation functions calculated with the QHNC compare very well for
the cases where AIMD simulations are available.  However, it turns out
that for most of the metals in our set a much simple linear response
formalism also performs remarkably well\cite{Loui98b}.  The surprising
accuracy of the linear response arises from a quantum interference
effect between two length-scales: the Fermi wave-vector and the
core-size of the ions, which makes the non-linear response terms much
less important than one might naively expect\cite{Loui98b}. The
accuracy of the QHNC also benefits from this effect.

The only metal in our set where the QHNC has difficulty is Ga.  We
found earlier\cite{Anta00} that an ad-hoc change in the local field
factor $G_{ee}(q)$ seems to improve matters.  We also found that the
d-electrons are close to a resonance.  These two points indicate that
approximation 1, discussed in the previous section, begins to break
down for Ga.

\section{Applying two-component DFT to inhomogeneous fluids}\label{section3}

In this section we explore the possibilities of studying inhomogeneous
metals within a two-component DFT formulation.  The advantage of
treating metals directly as as electron-ion mixtures in this way is
that we bypass the need for effective (density dependent) ion-ion
potentials.  Modern classical DFT methods use input from the
homogeneous liquid state\cite{Baus90,Evan92,Curt87,Dent89}; the QHNC
is ideally suited to provide such input for a two-component DFT.

One of the simplest approximations for the DFT of an inhomogeneous
system, first proposed by Ramakrishnan and
Yussouf(RY)\cite{Rama79,Haym81}, corresponds to truncating at second
order a functional Taylor expansion of the excess free-energy around
the homogeneous phase:
\begin{equation}\label{eq3.1}
\Delta F^{ex}[\rho_I,\rho_e] =- \sum_{\alpha} \sum_{\beta} 
\int d{\bf r} \int d{\bf r}' \Delta \rho_\alpha({\bf r}) C_{\alpha \beta} (\rho_\alpha^0, \rho_\beta^0; |{\bf r} - {\bf r}'|) \Delta  \rho_\beta({\bf r}'),
\end{equation}
where $\Delta F^{ex}= F^{ex}[\{\rho_\alpha({\bf r})\}] -
F_0^{ex}(\{\rho_\alpha^0\})$, and the direct correlation functions
$C_{\alpha \beta}(\rho_\alpha^0, \rho_\beta^0; |{\bf r} - {\bf r}'|)$
are those of the homogeneous liquid phase.  For metals, the QHNC
provides $C_{II}(r)$ and $C_{Ie}(r)$, while $C_{ee}(r)$ is fixed by
the jellium approximation.  Together, these correlation functions
completely determine the excess free-energy of Eq.(\ref{eq3.1}).

 To obtain the total free-energy, one then adds the ideal
contributions of the electrons and the ions.  For the ions, this ideal
free energy functional has the form:
\begin{equation}\label{eq3.2}
\beta F^{id}_I[\rho_I] = \int d{\bf r} \rho_I({\bf r}) \left\{ \ln
\left [ \rho_I({\bf r}) \Lambda ^3 \right] -1 \right\},
\end{equation}
but for the electrons the exact form is not known, and various
approximations must be made.  To preserve the advantages of the
current DFT approach, one needs a kinetic energy functional.  A number
of forms have been proposed; some are remarkably accurate, see
e.g.~\cite{Pear93} and references therein.  These kinetic energy
functionals also vary in the ease with which they can be implemented
in a full two-component DFT calculation\cite{Xu98}.

Classical DFT has frequently been applied to the freezing transition
and the fluid-solid interface of simple fluids\cite{Evan92}.  The
two-component DFT described above provides a basis upon which to build
similar applications for metallic systems.

As a first application, we attempted to use the two-component RY
formalism described above to calculate the freezing transition of Al.
A very similar approach was already implemented by one of
us\cite{Xu94} to study the freezing transition of H.  However there a
simpler semi-empirical prescription was used for $C_{Ie}(r)$.  Since
$H$ does not have a core radius, the cancellation of higher order
response terms found for simple metals\cite{Loui98b} does not hold.
This implies that the correlation functions and also the freezing
transition of $H$ should be more difficult to treat accurately than
would be the case for simple metals.  Other properties of liquid $H$,
such as the metal-insulator transition\cite{Nell98} are also much
harder to treat.  Partially for these reasons, we would initially
expect that the simple metals are actually better systems on which to
test rudimentary DFT's.

The one-body densities of the solid phase are parameterized by sums 
over the reciprocal lattice vectors (r.l.v.) $\{{\bf G}\}$:
\begin{equation}\label{eq3.3}
\rho_{\alpha}({\b r}) = \rho_\alpha^0 \sum_{\{G\}} \xi_\alpha^{G} \exp
\left[ i {\bf G}\cdot {\bf r}\right].
\end{equation}
For the ions a further, well established, simplification is to take
the real-space density $\rho_I({\bf r})$ as a sum of Gaussians,
centred on the lattice sites, which implies $\xi_I^{G} = \exp \left[
-G^2/4\varsigma\right]$.  With this density parameterization and
approximation, the full free-energy functional for the difference
between the solid and the liquid state free energy, taken from
Eqs.~\ref{eq1.1},~\ref{eq3.1} and~\ref{eq3.2}, reduces to:
\begin{eqnarray}\label{eq3.4}
\frac{\beta \Delta F}{N_I} = \left(\frac{3}{2} \ln \left[ \frac{\varsigma
d^2}{\pi} \right] - \frac{3}{2} - \ln 2\right) + \Delta F^{id}_e[\rho_e] & - &
\frac{1}{2} \sum_{\{{\bf G}\}} \,' (\xi_I^{{\bf G}})^2 \rho_I^0 \hat{c}_{II}(|{\bf
G}|) -
 \nonumber \\
\frac{1}{2} \sum_{\{{\bf G}\}} \,' (\xi_e^{{\bf G}})^2 \rho_e^0 \hat{c}_{ee}(|{\bf
G}|) & - & \sum_{\{{\bf G}\}} \,' \xi_I^{{\bf G}} \xi_e^{{\bf G}}
 \rho_e^0 \hat{c}_{Ie}(|{\bf G}|).
\end{eqnarray}
Here the $\hat{c}_{\alpha \beta}(|{\bf G}|)$ are the Fourier
transforms of the direct correlation functions for the liquid state,
and the primes mean that the term ${\bf G}=0$ should be omitted in the
sum.  The first term between brackets is the ideal free-energy of the
ions~(\ref{eq3.2}), written out in terms of the Gaussian parameter
$\varsigma$.

This free-energy can then be minimized through the variational
principle~(\ref{eq1.2}) to solve for the free-energy difference
between a given solid crystal structure and the fluid phase.  However,
when we attempted this for Al, with the direct correlation functions
given by the QHNC, we found that the free energy of the solid did not
have a stable minimum.  This is a common problem encountered when
applying a DFT to the freezing transition (see
e.g. ref~\cite{deKu90}): If the full direct correlation functions are
used, then it is very hard to stabilize solid phase.  If instead one
uses the DFT for a hard-sphere reference system, adding the attractive
interactions as a perturbation, very good agreement with experiment
and simulation is recovered.  Examples include phase
diagram\cite{Curt87} and the fluid-solid interface of the LJ
fluid\cite{Marr93}, and a one-component treatment of the freezing of
Al\cite{Dent99}, based on an effective pair potential.  This suggests
that a successful 2-component DFT treatment of the freezing transition
also needs to pass via this route.  Unfortunately this quickly reduces
to a theory similar to the one-component DFT, which means introducing
the effective pair potentials we are trying to avoid in the first
place.

Another application of the two-component DFT formalism is to examine
the fluid-solid interface of Al, for which Besson and
Madden\cite{Jess00} have recently completed extensive OF-AIMD
simulations.  They found non-trivial behaviour, with the ion-ion
density profiles decaying from an ordered structure to a smooth liquid
structure across several atomic layers.  Similarly, they found that
the valence electron density also showed oscillations which decayed
into the bulk liquid phase.  In principle, a two-component DFT would
be ideally suited to study this problem, since the ionic and
electronic density profiles come out on equal footing.  Unfortunately,
the difficulties encountered when trying to treat the freezing transition
also make studying this problem very difficult.  Instead, we tackle a
simpler problem: Given the ionic density profile, can we use our DFT
to calculate the electron density profile?

\begin{figure}
\begin{center}
\epsfig{figure=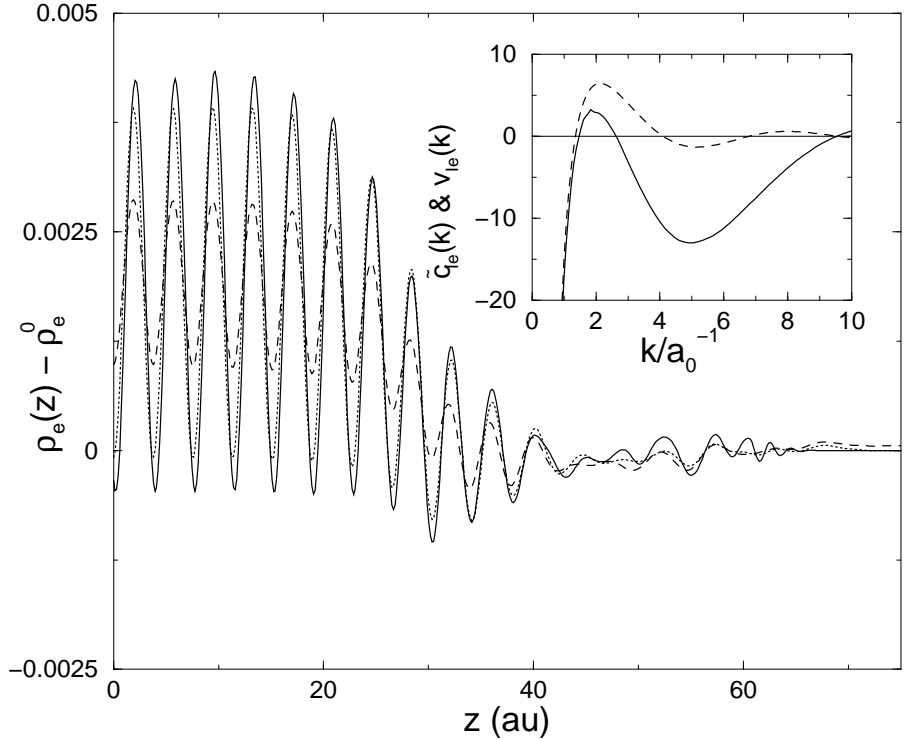,width=12cm} 
\begin{minipage}{14cm}
\caption{ Electron density profiles from the OF-AIMD
simulations\protect\cite{Jess00} (solid lines), compared to the QHNC-DFT
results (dashed lines) given by Eq.~(\protect\ref{eq3.7}), and linear
response theory (dotted lines) with an empty-core pseudopotential with
$R_c=1.153 a_0$.  Inset, comparison of $\hat{c}_{Ie}(k)$ (solid line)
and the pseudopotential $v_{Ie}(k)$ (dashed line), both in Hartrees.
\protect\label{fig:hong2}}
\end{minipage}
\end{center}
\end{figure}

If the ionic density profiles are given, which fixes the $\xi_I^{G}$,
then the electronic $\xi_e^{G}$ which satisfy the variational
principle~(\ref{eq1.2}) follow from Eq.~(\ref{eq3.4}):
\begin{equation}\label{eq3.7}
\xi_e^{{\bf G}} = -\frac{\xi_I^{{\bf G}} \hat{c}_{Ie}({\bf
G})}{\left(\beta/\chi^0_{ee}({\bf G})) + \hat{c}_{ee}({\bf G})\right)}.
\end{equation}  where we have made an additional approximation, also
made in \cite{Xu94}, namely that the ideal electron contribution
$\Delta F_e^{id}[\rho_e]$ is expanded to second order in $\xi_e^{\bf
G}$ (linear response).  This is consistent with the spirit of the RY
DFT approach, and greatly simplifies the minimization of the
electronic degrees of freedom, since they are now linearly coupled to
the ionic ones.  Note that taking the lowest order approximation to
the ion-electron direct correlation function, $C_{Ie}(r) \approx -
\beta v_{Ie}(r)$ in Eq.~(\ref{eq3.7}) would be equivalent to a
standard linear response calculation.  Since we use the full
$C_{Ie}(r)$, non-linear response terms are included, although one
would need to use the full kinetic energy functional to be completely
consistent with the RY DFT approach of Eq.~(\ref{eq3.4}).  The results
of using Eq.(\ref{eq3.7}) are shown in Fig.~\ref{fig:hong2} as a
function of the distance $z$ along the interface. (See
ref.~\cite{Jess00} for details of parameters).  The QHNC results are
in fact less accurate than a simpler linear-response approximation
which replaces $\hat{c}_{Ie}(G)$, with an empty-core Ashcroft
pseudopotential\cite{Ashc78,Hafn87}.  At first this may seem
surprising, but a careful look at the inset of Fig.~\ref{fig:hong2}
shows that the difference results from the fact that $\hat{c}_{Ie}(G)$
is considerably smaller at $G \approx 1.8$, where the strongest ionic
scattering occurs.  This difference stems in part from the fact that
the true pseudopotential in Al is known to be highly
non-local\cite{Hafn87} because of the $p$ character of the bonds,
suggesting that a local formulation, such as the simple linear
response DFT we employed, will not work well together with the QHNC
calculation which makes no local pseudopotential assumptions.
Non-linear response effects are also expected to be more important at
the highly inhomogeneous fluid-fluid interface.  In contrast to the
QHNC, the pseudo-potential contains an empirically adjustable
parameter which effectively includes non-local and non-linear effects.
To achieve accurate results with a DFT based on the QHNC, one needs to
treat the two approaches at comparable levels of approximation.  As is
often encountered in condensed-matter physics, mixing different levels
of approximation does not work very well.  Two ways of improving this
would be {\bf (1)} to use QHNC with the improved local
pseudopotentials used in the OFMD\cite{Jess00} or {\bf (2)} to use a
more sophisticated DFT approach.  Although far from complete or
completely satisfactory, this first, and rather primitive, example of
a two-component DFT used for a liquid-solid interface suggests that
our DFT approach could in principle be fruitfully used to study
fluid-solid interfaces in simple metals.

\section{Conclusions}

In conclusion, we have shown how DFT provides a unified formalism from
which to derive the QOZ equations, and the related QHNC approach,
first pioneered by Chihara.  The QHNC reduces the many-centre
electronic problem to an effective one-centre one.  This dramatically
reduces the computational time needed to calculate ion-electron and
ion-ion correlation functions.  It also has the advantage that no
explicit use of effective pair potentials or pseudo-potentials is
needed.  The most important approximation in the QHNC approach is to
treat the electron-electron correlations as those of jellium.  It is
not yet clear under which conditions this approximation begins to
break down.  With this caveat in mind, the other approximations
entering the QHNC are reasonably well understood.  We used the QHNC to
calculate the ion-electron radial distribution functions for a number
of simple metals, finding very good agreement with ab-initio molecular
dynamics calculations where these are available.

We have also discussed an exploratory application of the simple RY DFT
approach to liquid metals as two-component electron-ion mixtures.  We
found that describing the freezing transition suffers from a similar
problem to that found for classical simple fluids: the solid phase
does not develop a stable minimum if the full direct correlation
functions are used as input.  This is rather disappointing.  Slightly
more (partial) success was found when we attempted to use the DFT to
describe the inhomogeneous electron density at a liquid-solid Al
interface.  Although this result was only a partial application of the
DFT, since the ionic density profile came from the simulations, it
does suggest that developing a full fledged DFT could be very
fruitful.  Finally, it is clear that our rudimentary DFT approach is
not yet accurate enough, and could be improved in a number of ways. We
are attempting to generalize more sophisticated DFT schemes like the
MDWA\cite{Dent89} or GELA\cite{Baus90} to the two-component
ion-electron problem.  We are also studying the effect of using
different local field factors and different kinetic energy
functionals.

\acknowledgements
AAL acknowledges support from the Isaac Newton Trust, Cambridge, and
the hospitality of Lyd\'{e}ric Bocquet at the Ecole Normale Superieure 
in Lyon, where some of this work was completed.  He thanks
N.W. Ashcroft, P. Madden, and  M. Sprik for illuminating discussions.


\begin{thebibliography}{99}
\bibitem{core-electrons}In principle one could also define $g_{Ie}(r)$
for all the electrons, including the core-electrons, but since the
core electron density changes very little, they are not plotted for
the sake of clarity).
\end{thebibliography}
\end{document}